\documentclass[letterpaper]{article} 
\usepackage[preprint]{aaai2027}  
\usepackage[hyphens]{url}  
\usepackage{graphicx} 
\usepackage{natbib}  
\usepackage{caption} 
\usepackage{algorithm}
\usepackage{algorithmic}

\usepackage{newfloat}
\usepackage{listings}
\DeclareCaptionStyle{ruled}{labelfont=normalfont,labelsep=colon,strut=off} 
\floatstyle{ruled}
\newfloat{listing}{tb}{lst}{}
\floatname{listing}{Listing}

\usepackage{amsmath}
\usepackage{amssymb}

\usepackage{booktabs}

\title{Repository-Context-Guided Image Understanding for Agentic Software Development}
\author{
    Fumiki Yohena\textsuperscript{\rm 1},
    Toshiaki Sota\textsuperscript{\rm 1},
    Shota Suzuki\textsuperscript{\rm 1},
    Tomoko Okamoto\textsuperscript{\rm 1},
    Michiaki Tatsubori\textsuperscript{\rm 2}
}
\affiliations{
    \textsuperscript{\rm 1}IBM Japan Systems Engineering\\
    \textsuperscript{\rm 2}IBM Research
}

\begin{document}

\maketitle

\begin{abstract}
Coding agents increasingly receive UI screenshots as part of bug reports, yet degraded image quality—caused by low resolution, compression artifacts, or small rendered text—can prevent them from identifying the relevant UI elements and source files. 
Vision-language models become less reliable as image quality deteriorates, while OCR-assisted approaches may propagate recognition errors from degraded inputs. Both approaches are limited when the visual evidence required for localization is no longer clearly observable in the query screenshot. 
We propose augmenting coding agents with repository-rendered reference images: crisp screenshots generated from the same codebase that the agent is tasked with modifying. 
These references provide readable visual evidence of the application’s UI while preserving the current, potentially bug-containing implementation state. 
Because they are generated from the available repository rather than requiring a known bug-free version, they can be produced automatically for reproducible application states without manual image annotation. 
We evaluate the approach on a controlled benchmark derived from SWE-bench Multimodal by injecting visual UI bugs into a real web application.
We vary screenshot degradation and reference-image provision and measure whether the coding agent identifies the correct source file.
The results show that reference images improve file localization especially for degraded image quality, increasing the gold file match rate under fourfold downsampling with JPEG compression.
We confirmed with further analysis that the proposed approach recognizes text approximately 12 percentage points better than language biased OCR under the same degradation.
\end{abstract}


\section{Introduction}

Coding agents are increasingly applied to software engineering tasks that involve visual
inputs, such as GitHub issues with attached screenshots, UI bug reports, and feature
requests referencing specific interface elements~\cite{zhao2026beyond,huang2025guirepair}.
Benchmarks such as SWE-bench Multimodal~\cite{yang2024swebenchm} reflect this trend, demonstrating that a
meaningful fraction of real-world software issues require agents to interpret visual
attachments to identify the correct files and produce appropriate patches.
In practice, however, the screenshots available to agents can be degraded: images
attached to issue trackers may have undergone aggressive compression and software
under development may render UI text at sizes that fall below reliable recognition
thresholds.
When an agent misreads a UI label under such conditions, it queries irrelevant files or
components and fails to complete the task.

Existing systems handle degraded visual inputs in three main ways.
\textit{VLM-only} methods provide only the degraded screenshot to the VLM used
by the coding agent and support overall screenshot comprehension, but
their text-recognition performance can degrade for small UI text or low-quality screenshots~\cite{liu2024ocrbench,fu2025ocrbenchv2}.
\textit{OCR-assisted} pipelines supply OCR output to the model alongside the image~\cite{fhima2025tapvl}, but
OCR errors on degraded images propagate into downstream reasoning~\cite{singh2021textocr,biten2022latr}.
\textit{Lexicon-guided OCR} constrains recognition to a predefined vocabulary~\cite{nguyen2021dictionary,wei2023sitm}; however,
general-purpose lexicons lack coverage of application-specific strings, and methods that
derive candidates from the initial OCR result inherit its errors when degradation is
severe.

We observe that a source code repository is not merely a collection of text
files but also a renderable visual knowledge base: it can be used to 
programmatically produce high-quality screenshots of the application in
reproducible states.
Exploiting this observation, our method renders crisp reference images from the
repository and supplies the references alongside the degraded screenshot to a VLM.
The reference images provide readable visual evidence that compensates for
information lost through degradation, without requiring any text-level
extraction from the degraded input.
We evaluate our approach on a controlled benchmark built on a real web
application repository included in SWE-bench
Multimodal.
The benchmark comprises 36 injected visual UI bugs and evaluates source-file
localization and UI-text recognition under varying screenshot-degradation and
reference-image provision conditions.
Under the most severe degradation condition, repository-rendered reference
images increase the gold-file match rate for localizing the bug-causing source
file from 28\% to 42\%.
We make two contributions:
\begin{itemize}
    \item We propose a method that augments coding agents with
          repository-rendered reference images: the repository is used as a
          visual knowledge base from which crisp screenshots are generated and provided as visual context for interpreting degraded query images.
    \item We construct a controlled benchmark by injecting 36 visual UI bugs
          into a real web application included in SWE-bench Multimodal, and evaluate source-file localization
          and UI-text recognition under varying screenshot-degradation and reference-provision conditions.
\end{itemize}

\section{Related Work}

This section reviews related work from three perspectives: coding agents that operate on visual 
inputs, GUI understanding methods that parse screenshots into structured interface elements, and scene-text
recognition methods, including OCR-assisted and lexicon-guided approaches.

Recent work has extended coding agents to repository-level software
engineering tasks involving screenshots and other visual artifacts
~\cite{zhao2026beyond,huang2025guirepair}.
SWE-bench Multimodal formalizes such tasks through real-world GitHub issues
whose resolution depends on visual attachments
~\cite{yang2024swebenchm}.
These studies establish the importance of visual inputs in software issue
resolution, but focus primarily on interpreting the supplied images rather
than recovering visual information lost through degradation.

Methods such as ScreenAI~\cite{baechler2024screenai} and
OmniParser~\cite{lu2024omniparser} analyze screenshots into UI
elements---buttons, input fields, menus, icons, and text regions---providing a
foundation for GUI agents to recognize and act on screen
content~\cite{li2023spotlight,cheng2024seeclick,song2024visiontasker,niu2024screenagent,you2024ferretui,he2024webvoyager}.
This line of GUI understanding shares with our work the view of screenshots as
structured interfaces rather than raw images.
In our approach, UI-element descriptions can serve as retrieval cues for
selecting relevant repository-rendered reference images, rather than as the
final output.

Scene text recognition has long been studied as a distinct challenge from
document OCR~\cite{jung2004text}.
Text in natural images is affected by background clutter, lighting variation,
perspective distortion, blur, and diverse fonts and layouts, and prior work has
shown the importance of integrating detection, binarization, and recognition
into unified pipelines~\cite{weinman2014integrated,ye2015survey}.
Transformer-based OCR models have advanced
recognition quality on printed and scene text~\cite{li2022trocr,atienza2021vitstr,bautista2022parseq}, while recent studies on the OCR
capabilities of multimodal large language models show competitive performance
under favorable conditions but continued sensitivity to image
resolution~\cite{inoue2025context,liu2024ocrbench,fu2025ocrbenchv2}.
Another line of work augments vision-language models with text and layout
information extracted by external OCR systems, but errors in the OCR output
can limit downstream reasoning~\cite{fhima2025tapvl,singh2021textocr,biten2022latr}.
The effectiveness of constraining the candidate space has been demonstrated by
lexicon-guided and dictionary-guided scene text recognition, which improve
accuracy by using dictionary candidates to guide recognition~\cite{nguyen2021dictionary,wei2023sitm}.
Dynamic Lexicon Generation~\cite{patel2016dynamic} further shows that generating
word candidates from the visual context of the whole image can reduce the
candidate space.
The candidate sources used in existing lexicon-guided methods are primarily
general-purpose dictionaries, image-level scene context, or initial OCR
results~\cite{patel2016dynamic,nguyen2021dictionary,wei2023sitm}.
General dictionaries may be too broad relative to the specific strings in a
given image, and methods that rely on initial recognition results remain vulnerable when those
results are severely corrupted by image degradation.

To our knowledge, prior work has not treated the current source repository as
a renderable visual knowledge base for interpreting degraded issue screenshots.
Our approach generates high-quality reference images from the
current repository and provides them alongside the degraded screenshot.
These references supply the coding agent with readable visual evidence for UI-text
recognition and downstream source-file localization, without requiring an
external OCR transcription of the degraded input.

\section{Problem Setting}

This section formalizes image understanding for source-file localization from
degraded UI screenshots.
We first describe the ordinary localization workflow, then characterize how
image degradation disrupts visual understanding, and finally formulate our
repository-context-guided approach using repository-rendered reference images.

\begin{figure*}[t]
  \centering
  \includegraphics[width=\textwidth]{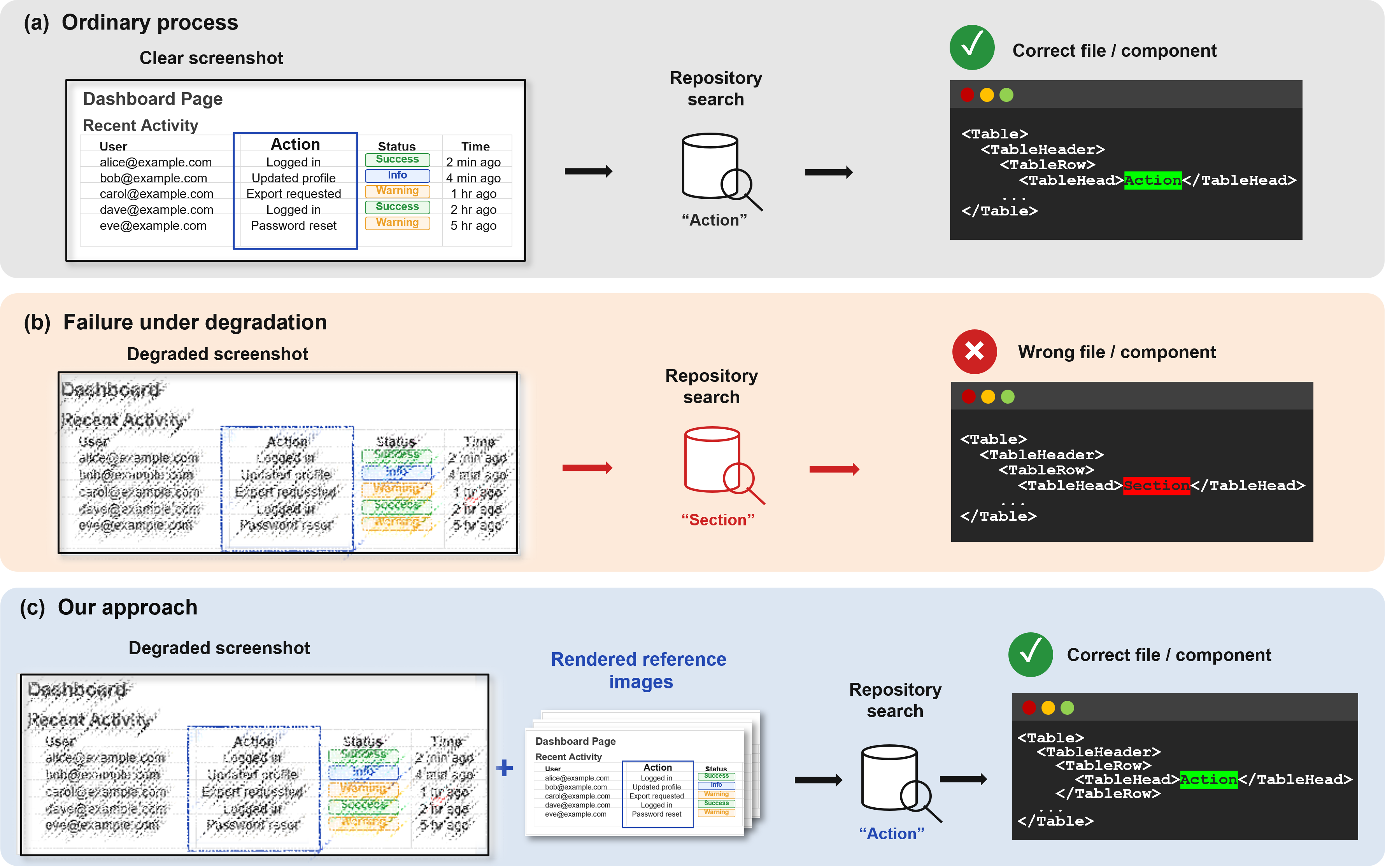}
  \caption{Problem setting for source-file localization from degraded UI screenshots.
    (a)~The coding agent extracts visual evidence from a clear screenshot and
    retrieves the correct source file.
    (b)~Image degradation can corrupt visual evidence and misdirect repository search, resulting in incorrect file localization.
    (c)~Repository-rendered reference images provide additional visual evidence that supports reliable interpretation of the degraded screenshot and source-file localization.}
  \label{fig:problem}
\end{figure*}

\paragraph{Ordinary workflow.}
Let $\mathcal{R}$ denote a source code repository, $d$ a task description,
and $I$ a clear UI screenshot.
The coding agent first uses a VLM $f$ to extract visual evidence
\begin{equation}
  \hat{Z} = f(d,\, I),
  \label{eq:workflow}
\end{equation}
where $\hat{Z}$ may include recognized text, UI elements, and spatial cues.
The agent uses this evidence to search the repository and predict a set of
relevant source files (Figure~\ref{fig:problem}(a)):
\begin{equation*}
  \hat{\mathcal{F}} = s(\mathcal{R},\, d,\, \hat{Z}),
  \label{eq:file_localization}
\end{equation*}
where $s$ denotes the repository-search and file-localization process.
UI text can provide a useful search cue because labels visible in the
screenshot may also appear in component definitions or related source files.

\paragraph{Failure under degradation.}
In practice, screenshots attached to bug reports can be degraded by
low resolution, compression, or rescaling.
Let $I' = g(I)$ denote such a degraded observation, where $g$ denotes a
degradation process that is not explicitly provided to the agent.
The visual evidence extracted from $I'$ may be unreliable; in particular, recognized UI 
text and other visual cues may be corrupted, causing the agent to
retrieve incorrect files (Figure~\ref{fig:problem}(b)).

\paragraph{Our approach.}
We compensate for information lost during degradation using reference images
rendered from the current repository.
Let $\mathcal{C}_{\mathcal{R}}$ denote the set of repository-rendered
reference images.
The VLM uses the task description $d$, the degraded screenshot $I'$, and the
repository-rendered reference images $\mathcal{C}_{\mathcal{R}}$ to produce
recovered visual evidence:
\begin{equation*}
  \hat{Z}_{\mathrm{ref}}
  = f(d,\, I',\, \mathcal{C}_{\mathcal{R}}).
\end{equation*}
As with $\hat{Z}$ in Equation~\eqref{eq:workflow},
$\hat{Z}_{\mathrm{ref}}$ may include recognized UI text, UI elements, and
spatial cues.
The coding agent then uses this evidence to localize relevant source files (Figure~\ref{fig:problem}(c)):
\begin{equation*}
  \hat{\mathcal{F}}_{\mathrm{ref}}
  = s(\mathcal{R},\, d,\, \hat{Z}_{\mathrm{ref}}).
  \label{eq:reference_localization}
\end{equation*}

\section{Method}

\begin{figure*}[t]
  \centering
  \includegraphics[width=\textwidth]{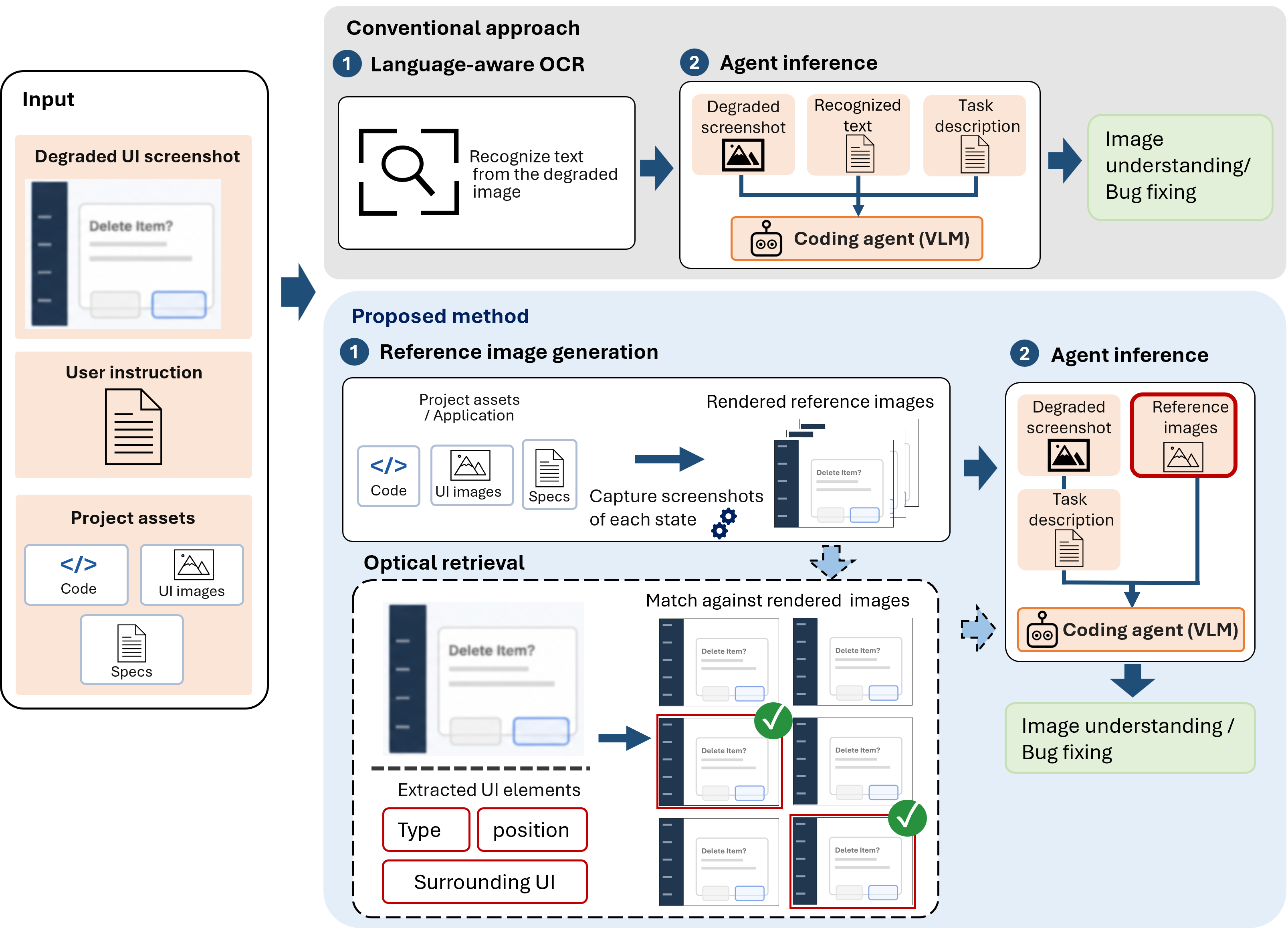}
  \caption{Overview of the proposed method.
An OCR-assisted approach first extracts text from the degraded screenshot and
provides the transcription alongside the image to the coding agent.
Our method instead renders high-quality reference images from the current
repository and provides them together with the degraded screenshot and task
description.
For larger reference catalogs, optional retrieval selects a subset of relevant references.}
  \label{fig:method}
\end{figure*}

The core idea of our method is to render high-quality reference images from
the current repository revision and provide them as visual context for
interpreting degraded screenshots.
Rather than relying solely on visual evidence in the degraded query, the coding agent
jointly observes the query and the repository-rendered references.
The references preserve readable text, layout, and surrounding UI structure,
supporting more reliable image understanding
(Figure~\ref{fig:method}).
The proposed method is detailed below.

\subsection{Reference Image Generation}

We build the target repository and render a predefined set of reproducible
application states to obtain high-quality screenshots.
Specifically, we launch the target application and use Playwright
\footnote{\url{https://playwright.dev}} to navigate to a predefined set of
reproducible UI states and capture screenshots of each state.
Each screenshot is associated with its UI-state identifier, forming a catalog
of the application states covered by the capture script.

When the reference catalog is sufficiently compact, we provide all reference
images directly to the coding agent.
When providing the entire catalog is impractical because of input size or
inference cost, we optionally select a subset of relevant references.
Let $D(I')$ denote the textual UI-structure description obtained by parsing
the degraded screenshot $I'$.
We embed the screenshot and its description into a shared multimodal embedding
space:
\begin{equation*}
  \mathbf{q}_{\mathrm{img}} = e_{\mathrm{img}}(I'),
  \qquad
  \mathbf{q}_{\mathrm{txt}} = e_{\mathrm{txt}}(D(I')),
\end{equation*}
where $e_{\mathrm{img}}$ and $e_{\mathrm{txt}}$ denote the image and text
encoders of a shared multimodal embedding model, respectively.
In our experiments, we use Gemini Embedding~2%
\footnote{\url{https://ai.google.dev/gemini-api/docs/models/gemini-embedding-2}}
as the multimodal embedding model.
When a usable UI-structure description is available, the retrieval query is
formed by element-wise averaging:
\begin{equation*}
  \mathbf{q}
  =
  \frac{1}{2}
  \left(
    \mathbf{q}_{\mathrm{img}}
    +
    \mathbf{q}_{\mathrm{txt}}
  \right).
\end{equation*}
Otherwise, we use $\mathbf{q}=\mathbf{q}_{\mathrm{img}}$.
For each reference image
$C_j \in \mathcal{C}_{\mathcal{R}}$, we pre-compute
$\mathbf{c}_j=e_{\mathrm{img}}(C_j)$ and rank the catalog according to
\begin{equation*}
  s_j = \cos(\mathbf{q},\,\mathbf{c}_j),
\end{equation*}
where $j$ indexes the reference images.
The top-$k$ reference images with the highest similarity scores are then
provided to the coding agent.

\subsection{Reference-Augmented Agent Inference}

We provide the degraded screenshot and the repository-rendered reference
images together to the coding agent for UI understanding.
The coding agent leverages the clearly rendered content in the reference images as cues
to interpret unclear elements in the degraded image.
The coding agent receives the task description, the degraded screenshot, and
high-quality repository-rendered reference images as input. It then uses this
augmented visual context for source-file localization and patch generation.

The distinguishing characteristic of the proposed method is that it provides
repository-derived evidence at the image level rather than the text level.
The OCR-based baselines supply the model with recognized strings---optionally
augmented by a repository-derived lexicon---but carry no information about where
on the screen each string appears or what the surrounding visual context looks
like.
In contrast, rendered reference images preserve the full visual
context---including text placement, font, color, and surrounding UI
structure---allowing the coding agent to establish spatial correspondences for recognition.

\section{Experiments}

\begin{figure*}[t!]
  \centering
  \includegraphics[width=\textwidth]{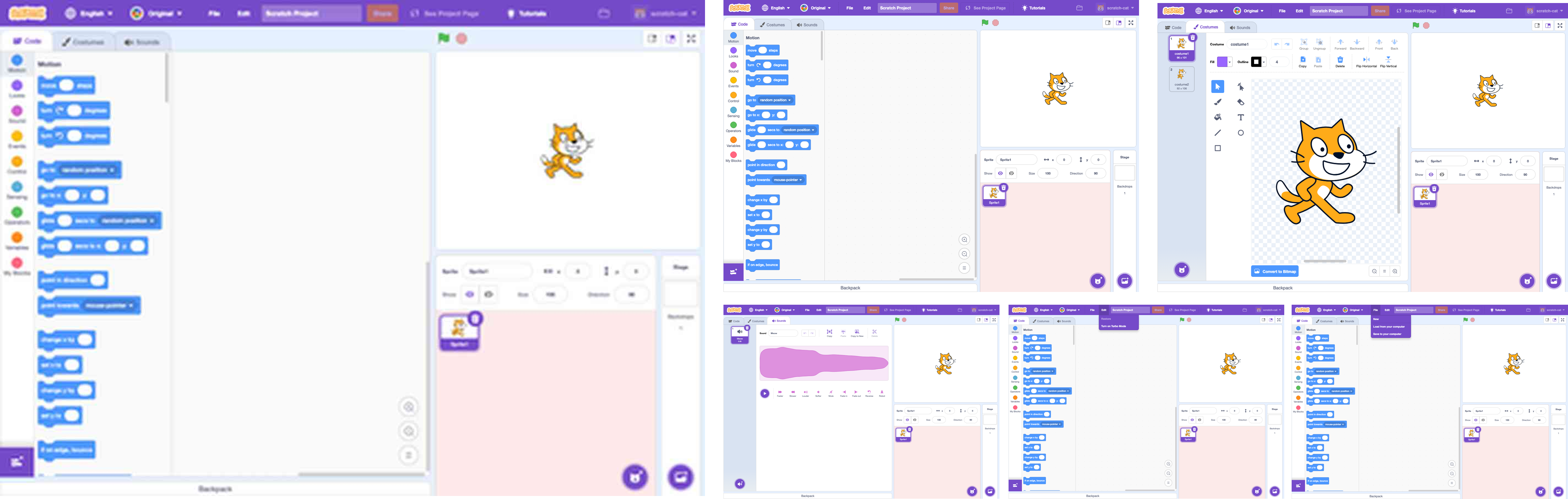}
  \caption{Example task instance (CSS color mutator).
The image on the left is the degraded query screenshot (320\,px), while the five images on the right comprise the reference catalog
rendered from the bug-injected codebase via Playwright.}
  \label{fig:dataset-example}
\end{figure*}

We evaluate whether providing repository-rendered reference images improves
the coding agent's ability to localize the responsible source file for visual UI
bugs when the issue screenshot is degraded.
We inject 36 visual bugs into a real application drawn from
SWE-bench Multimodal and conduct three sets of
experiments under controlled conditions:
(1)~measuring gold file match across image resolutions and reference-provision
settings;
(2)~isolating text recognition accuracy to examine how degradation affects
the readability of on-screen text, and whether reference images compensate
for this loss;
(3)~varying the specificity of issue descriptions to delineate when visual
context is beneficial versus redundant.

\subsection{Dataset Construction and Image Preparation}

\paragraph{Data source.}
We construct task instances on scratchfoundation/scratch-gui\footnote{\url{https://github.com/scratchfoundation/scratch-gui}} (Scratch~3.0
editor, React + CSS Modules), one of the repositories in
SWE-bench Multimodal, inspired by the mutation-based
instance-generation methodology of SWE-smith~\cite{yang2025swesmith}.
We implement six deterministic mutators (text typo, text mismatch, icon swap,
CSS flex-direction, CSS size, CSS color) and apply them to produce candidates.
Each bug modifies one source file.
We retain the 36 instances whose pixel difference from the clean state falls
within $[0.002\%,12\%]$ to focus on bugs that produce visible but spatially limited changes.
The dataset contains 7 text-typo, 7 text-mismatch, 10 icon-swap,
6 CSS flex-direction, 3 CSS-size, and 3 CSS-color instances.
Pixel difference is measured as the fraction of pixels whose maximum absolute
RGB-channel difference between the clean and buggy screenshots exceeds 12.
Issue text describes only the symptom category (e.g., ``one of the text labels
seems wrong'') and references the screenshot, with no file or component names.

\paragraph{Query and reference image preparation.}
For each of the 36 bug instances, we capture an undegraded query screenshot at a
$1280 \times 900$ viewport.
We generate degraded queries by resizing the screenshot to widths of 320, 480,
or 640\,px, upscaling it back to the original dimensions using Lanczos
interpolation, and saving it as JPEG at quality 30.
For reference image preparation, we use Playwright to render the application
from the current bug-injected codebase and obtain a catalog of five crisp
screenshots corresponding to the five predefined states.
Figure~\ref{fig:dataset-example} shows an example of a degraded query and its corresponding reference catalog.

\subsection{Effect of Reference Image Provision}
\label{sec:main-results}
For reference provision, we compare three settings: (1)~no reference images
(None), (2)~retrieval top~1, where a single image is selected from the
catalog via Gemini Embedding~2 cosine similarity (Top~1 reference), and (3)~the
full catalog of 5 crisp screenshots rendered from the current
(bug-injected) codebase (All references).
For the Top~1 reference condition, the retrieval query combines the image
embedding with a textual UI-element description produced by OmniParser at
640 and 480\,px.
At 320\,px, OmniParser produces no detections for 34 of the 36 queries.
We therefore use image-only retrieval for all 36 queries at this resolution
to maintain a consistent retrieval setting.
As the coding-agent framework, we use mini-swe-agent,\footnote{\url{https://github.com/SWE-agent/mini-swe-agent}} a publicly available lightweight coding-agent
framework from the SWE-agent project~\cite{yang2024sweagent}.
Each instance is solved inside a Docker container; the agent views the issue
screenshot via the \texttt{view\_image} command and the agent is not informed of the degradation.
The LLM backend is GPT-5.4 with temperature 0.
Each condition is evaluated once, and the agent is limited to 75 steps per
instance.
Bug patches are applied at container startup and \texttt{.git} is
re-initialized to prevent history-based shortcuts.
We report the gold-file match rate, defined as whether the agent modifies
the single file changed by the gold patch and no additional files.
This directly measures localization accuracy independent of patch correctness.

\begin{table}[t]
\centering
\begin{tabular}{lccc}
\toprule
Query & None & Top~1 reference & All references \\
\midrule
Undegraded   & 33 & --- & --- \\
640\,px & 28 & 28  & \textbf{36} \\
480\,px & 22 & 22  & \textbf{36} \\
320\,px & 28 & 28  & \textbf{42} \\
\bottomrule
\end{tabular}
\caption{Gold file match (\%) across query degradation levels and reference
provision settings. Top~1 reference selects one image from the
catalog via embedding similarity; All references provides all five images.}
\label{tab:main-results}
\end{table}

Table~\ref{tab:main-results} shows gold file match across query degradation
levels and reference provision settings.
Without reference images, degradation reduces performance from 33\% (Undegraded) to
22--28\% across degraded conditions.
The Top~1 reference condition selects the single most similar reference image
using Gemini Embedding~2, but achieves the same performance as the
no-reference condition at all degradation levels.
In contrast, providing the full catalog of 5 images consistently improves
performance: +8\,pp at 640\,px, +14\,pp at 480\,px, and +14\,pp at
320\,px, exceeding the undegraded no-reference baseline in all degraded conditions.

\subsection{Text Recognition Under Degradation}
\label{sec:textrecog}

\begin{figure*}[t!]
\centering
\includegraphics[width=\textwidth]{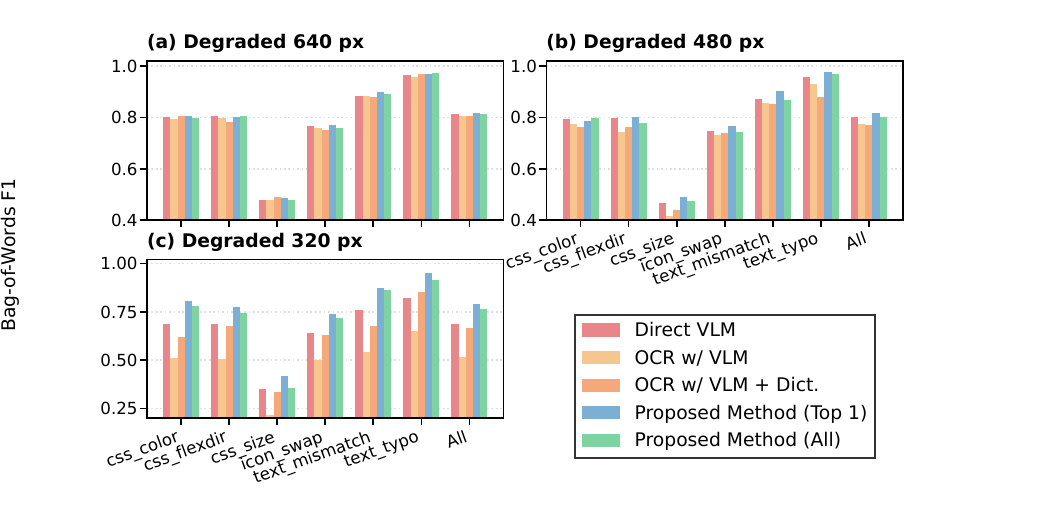}
\caption{Mean BoW-F1 by bug type at
  (a)~640\,px, (b)~480\,px, and (c)~320\,px.
  The reference-augmented methods increasingly outperform the OCR-based
  baselines as degradation becomes more severe.}
\label{fig:textrecog}
\end{figure*}

The preceding experiment measures whether agents can localize the
responsible source file; here we isolate the more basic question of whether
the degraded screenshot's UI text can even be read.
We define Bag-of-Words F1 (BoW-F1), adapting prior
order-invariant bag-of-words evaluation for text recognition
~\cite{vidal2023endtoend}, as follows: each text is
tokenized into lowercased alphanumeric tokens of length $\geq 2$ and
represented as a multiset (i.e., duplicate tokens are counted).
Let $P$ and $G$ denote the predicted and ground-truth token multisets.
True positives are computed as
\[
  \mathrm{TP} = \sum_{w} \min\bigl(P(w),\, G(w)\bigr)\;,
\]
where $P(w)$ and $G(w)$ denote the multiplicities of token $w$ in the respective
multisets.
Precision and recall are $\mathrm{TP}/|P|$ and $\mathrm{TP}/|G|$
respectively, and F1 is their harmonic mean.
Ground-truth text is obtained by having GPT-5.4 read the undegraded
$1280\times900$ screenshot, then manually verified and corrected by the
authors.
For each resolution, we report BoW-F1 macro-averaged over the 36 instances.

We compare five approaches:
Direct VLM (the degraded image is passed directly to VLM without
additional context),
OCR w/ VLM (the degraded image and text extracted by Azure AI Document Intelligence%
\footnote{\url{https://learn.microsoft.com/en-us/azure/ai-services/document-intelligence/overview}} using the \texttt{prebuilt-read} model are passed to VLM for
correction),
OCR w/ VLM + Dict.\ (the degraded image, OCR output, and a 761-entry vocabulary extracted from
the repository's \texttt{scratch-l10n} locale files are passed to VLM),
and the Proposed method, evaluated under two reference-provision settings:
Top~1 reference, which provides the highest-ranked reference image, and
All references, which provides all five reference images alongside the
degraded image.
All model-based approaches use GPT-5.4.
The Top~1 reference condition follows the same retrieval protocol as in the preceding 
source-file localization experiment, using image and OmniParser-derived textual embeddings 
at 640 and 480\,px, and image-only retrieval at 320\,px where OmniParser fails to produce detections.

When the undegraded 1280$\times$900 screenshot is used as input, all methods
achieve F1 = 1.000, indicating that the task difficulty stems from
resolution degradation.
Figure~\ref{fig:textrecog} shows the results.
At degraded 640\,px (Figure~\ref{fig:textrecog}(a)), all five methods perform
comparably (F1 $\approx$ 0.81--0.82), with Proposed Method (Top~1) achieving the
highest overall score (0.819) and Proposed Method (All) scoring 0.815.
At degraded 480\,px (Figure~\ref{fig:textrecog}(b)), Proposed Method (Top~1) maintains its
lead (F1 = 0.818), while Proposed Method (All) scores 0.800.
OCR w/ VLM begins to degrade on \texttt{css\_size} instances (F1 = 0.42).
At degraded 320\,px (Figure~\ref{fig:textrecog}(c)), OCR w/ VLM degrades
substantially (F1 = 0.515) because the OCR engine struggles to extract text at
low resolution.
OCR w/ VLM + Dict.\ slightly underperforms Direct VLM (0.665 vs.\ 0.686),
indicating that dictionary-assisted correction cannot fully compensate for
poor OCR quality.
Proposed Method (Top~1) achieves the highest mean F1 (0.792, +0.105 over Direct VLM), while Proposed Method (All) also
outperforms all baselines (0.765, +0.079 over Direct VLM) but falls below
Top~1.
This suggests that the four non-matching reference images act as visual noise
that distracts the VLM from the relevant reference.

Notably, the opposite pattern holds for file localization (Table~\ref{tab:main-results}): retrieval 
top~1 provides no benefit, and the full catalog is required.
This asymmetry suggests that text recognition benefits from a focused, 
high-relevance reference, whereas file localization requires broader visual 
context that a single reference cannot supply.
Because our catalog contains only five images, providing the full catalog is practical. 
For larger catalogs, however, retrieval may be necessary to control input size 
while retaining sufficient visual context.

\subsection{Impact of Issue Description Granularity}
\label{sec:granularity}

To delineate when screenshot attachment is beneficial versus redundant, we
vary the specificity of the issue description along a five-level ladder
while keeping other conditions fixed: the same 36~instances, undegraded
(1280$\times$900) screenshots, the same agent and model (GPT-5.4 with
mini-swe-agent), and a single run per condition.
No reference image catalog is used; the ``with image'' condition simply
attaches the undegraded screenshot to the issue description body, while ``no
image'' replaces it with a notice that the link is unavailable.

Levels L0--L4 each cumulatively add information to the issue description.
The following shows the varying portion of each level for one example 
instance (text-label bug in the \texttt{menu-bar} component, Code tab
screen):
\begin{itemize}
  \item \textbf{L0} (generic description): ``Something in the UI doesn't look
        right.''
  \item \textbf{L1} (+ symptom category): adds ``\,---\,one of the text
        labels somewhere in the UI seems wrong.'' 
  \item \textbf{L2} (+ screen name): adds ``somewhere on the Code tab
        screen.''
  \item \textbf{L3} (+ component name): adds ``in the \texttt{menu-bar} area
        on the Code tab screen.''
  \item \textbf{L4} (+ specific values): adds ``a label that should read
        \texttt{"File"} is displayed as \texttt{"Flie"}.''
\end{itemize}
All prompts are generated deterministically from templates without LLM
involvement.
L4 serves as a sanity-check condition in which the target file is explicitly
identifiable from the description alone.

\begin{table}[t]
\centering
\setlength{\tabcolsep}{5pt}
\begin{tabular}{llccc}
\toprule
Level & Added content & No img & With img & $\Delta$ \\
\midrule
L0 & (generic description) &  3 &  9 & $+6$ \\
L1 & + symptom category &  9 & 12 & $+3$ \\
L2 & + screen name      &  9 & 14 & $+5$ \\
L3 & + component name   & 26 & 23 & $-3$ \\
L4 & + specific values  & 36 & 35 & $-1$ \\
\bottomrule
\end{tabular}
\caption{Gold file match (out of 36) by issue description level and
screenshot attachment.  $\Delta$ = (With img) $-$ (No img).}
\label{tab:granularity}
\end{table}

Table~\ref{tab:granularity} shows gold file match across description levels.
In the no-image condition, performance increases from 9/36 at L2 to 26/36
at L3.
The corresponding increase in the with-image condition is from 14/36 to
23/36, suggesting that component names provide strong cues for source-file
localization.
Below L3, attaching a screenshot consistently improves localization by $+3$
to $+6$ matches; at L3 and above, the text description already provides
sufficient localization information and attaching a screenshot yields no
positive marginal benefit ($\Delta \leq 0$).

\subsection{Limitations}

The current implementation relies on Playwright-based rendering scripts to
produce the reference catalog, limiting its direct applicability to web
applications with a working build pipeline.
Extending coverage to native desktop or mobile applications would require
alternative rendering toolchains.

Because the catalog is generated from the same codebase version as the
bug-injected build, reference images may themselves contain the target bug;
the method provides higher-fidelity views of the application state, not
necessarily bug-free ground truth.
Furthermore, the catalog covers only those screen states exercised by the
rendering scripts and cannot surface bugs that manifest only under transient
user interactions or runtime-dependent conditions.

Our benchmark comprises 36 synthetically injected bugs in a single React
application, evaluated with one coding agent and one VLM backend.
Image degradation is limited to controlled resizing with JPEG compression;
other real-world degradations such as camera capture, occlusion, or motion
blur are not studied.
The evaluation metric (gold file match) measures source-file localization
rather than patch correctness, and the reference catalog contains only five
images per instance, leaving the impact of catalog scale unexplored.

\section{Conclusion}

We presented a method that leverages repository-rendered reference images as
visual context for coding agents operating on degraded screenshots.
Supplying undegraded renderings alongside the degraded input provides a visual
prior that compensates for information lost through compression and resizing.
Our experiments show that repository-rendered reference images improve
source-file localization under degraded conditions, with the full catalog
improving the gold-file match rate.
They also improve UI-text recognition, where the Top~1 reference achieves the
best performance, particularly under severe degradation.
Future work includes extending the reference catalog to multi-state
renderings that capture interactive and runtime-dependent UI variations,
scaling to large repositories, and evaluating
generalization across diverse degradation conditions
beyond those studied here.

\bibliography{aaai2027}


\end{document}